\documentclass[aps,twocolumn]{revtex4-2}

\usepackage[utf8]{inputenc}
\usepackage{listings}

\usepackage{amsmath,bm}
\usepackage{amssymb}
\usepackage{graphicx}
\usepackage[table]{xcolor} 

\usepackage[english]{babel}
\usepackage{epsfig}
\usepackage{graphicx}
\usepackage{color}
\usepackage[normalem]{ulem}
\usepackage{url}
\usepackage{float}
\usepackage[breaklinks, plainpages=false, colorlinks=true, anchorcolor=cyan, linkcolor=red, citecolor=cyan, urlcolor=magenta, bookmarks=false]{hyperref}

\usepackage[caption=false]{subfig}

 \def\be{\begin{equation}}
\def\ee{\end{equation}}
 \def\ba{\begin{align}}
\def\ea{\end{align}}
\def\bea{\begin{eqnarray}}
\def\eea{\end{eqnarray}}

\def\th{{\bm{\Vec{\theta}}}}

\def\BW{{\tt BayesWave }}
\def\BWV{{\tt BayesWaveVoices }}

\newcommand{\bseq}{\begin{subequations}}
\newcommand{\eseq}{\end{subequations}}

\begin{document}

\title{Detecting gravitational wave signals using a flexible model for the amplitude and frequency evolution}

\author{Toral Gupta}
\affiliation{eXtreme Gravity Institute, Department of Physics, Montana State University, Bozeman, Montana 59717, USA}

\author{Neil J. Cornish}
\affiliation{eXtreme Gravity Institute, Department of Physics, Montana State University, Bozeman, Montana 59717, USA}

\email{toralgupta@montana.edu}

\date{\today}

\begin{abstract}
We currently lack good waveform models for many gravitational wave sources. Examples where models are lacking include neutron star post merger signals, core collapse supernovae, and signals of unknown origin. Wavelet based techniques have proven effective at detecting and characterizing these signals. Here we introduce a new method that uses collections of evolving amplitude-frequency tracks, or ``voices'', to model generic gravitational wave signals. The analysis is implemented using trans-dimensional Bayesian inference, building on the earlier wavelet-based \BW algorithm. The new algorithm, {\tt BayesWaveVoices}, outperforms the original for long duration signals.
\end{abstract} 
\vspace{-1cm}

\maketitle
\section{Introduction}\label{sec:intro}
Black holes and neutron stars stand out as the two most prominent sources of gravitational Waves (GW) observed to date~\cite{PhysRevX.9.031040,Abbott_2021,Abbott_2023} using ground-based detectors such as the Laser Interferometer Gravitational-Wave Observatory (LIGO) in the USA, the Virgo detector in Italy~\cite{Aasi_2015,PhysRevD.102.062003,Acernese_2015,PhysRevLett.123.231108} and the Kagra detector in Japan~\cite{10.1093/ptep/ptaa125}. With the increasing number of detections and improved detector sensitivity, the observation of longer-duration signals has become possible. This emphasizes the necessity for robust signal models capable of capturing signatures over extended time spans. While models for compact binary mergers have steadily improved in the last two decades, we currently lack reliable models for many interesting sources, such as the post merger dynamics of binary neutron stars and core collapse supernovae. Additionally, there may be signals coming from unanticipated sources, for which no models exist. And while general relativity has successfully passed all current tests, future observation of longer duration signals, with high signal to noise, will allow for more stringent tests of the theory. These considerations motivate the development of techniques that can detect and reconstruct arbitrary gravitational wave signals.

Ground-based detectors are most sensitive to transient signals from Compact Binary Coalescence (CBC) systems, particularly those involving Neutron Stars and Black Holes, with a total mass $M_T$ falling within the stellar mass range ($2M_\odot \leq M_T \leq 200 M_\odot$). The gravitational wave emission from these systems become most pronounced in the sensitive frequency band of the LIGO-Virgo-Kagra detectors (20 - 1000 Hz) ~\cite{Harry_2010,Ghonge:2020suv}, reaching a peak amplitude milliseconds before the merger. The these observations allows for the potential to uncover new physical effects in the strong gravity regime.

Unlike traditional match-filtering techniques~\cite{PhysRevD.85.122006} that rely on accurate template waveforms derived from theoretical models and numerical simulations~\cite{PhysRevLett.96.111101, PhysRevLett.96.111102, PhysRevLett.95.121101}, wavelet based methods have been developed to detect and characterize generic signals. Prominent examples include coherent Wave Burst~\cite{Klimenko:2005xv,Klimenko_2016,drago2021coherent} and a Bayesian algorithm known as \BW~\cite{Cornish_2015, Cornish_2021}. These algorithms stand apart by avoiding assumptions of a predefined template for signal detection, instead utilizing a wavelet basis, or frame, that allows the data to determine the possible morphology of the signal. \BW employs the sum of continuous Morlet-Gabor wavelets to reconstruct a signal, where both the number of wavelets and the parameters of the wavelets are explored by a trans-dimensional sampling algorithm. \BW performs very well on short duration signals, but the performance drops for long duration signals. Specialized search techniques have been developed for long duration burst signals, including seedless clustering algorithms~\cite{Thrane:2010ri,Thrane:2013bea,Thrane_2014,PhysRevLett.115.181102} and variants of the coherent Wave Burst algorithm that are tuned to detect longer duration signals~\cite{Klimenko_2016}. However, there are currently no effective Bayesian inference algorithms for reconstructing long duration burst signals.

To address these challenges, a new algorithm, {\tt BayesWaveVoices}, is introduced in this study. The name is inspired by the literature on extreme mass ratio inspirals, where the signal can be decomoposed as a collection of ``voices'', each with a define amplitude and frequency evolution~\cite{Hughes:2021exa}. These ``voices'' do not necessarily have instantaneous frequencies that are related by integer multiples, so the term ``voices'' is used instead of ``harmonics''. For un-expected or un-modeled sources, we have reason to expect that the gravitational waves they produce can be described by a collection of voices with time evolving amplitudes and frequencies. Gravitational waves are sourced by time varying stress-energy and current multipole moments, with bulk, coherent motion likely to produce the strongest contribution~\cite{poisson_will_2014,Blanchet_2014}. 
This is certainly true of known sources such as compact binaries, including those beyond GR, but it will also be true for any source where the emission is dominated by a few multipoles~\cite{Kastha_2018,PhysRevD.100.044007}. This reasoning extends to more complicated examples such as the post-merger dynamics of neutron star binaries~\cite{Clark_2016}. Numerical simulations of astrophysical sources such as core collapse supernovae show evidence for a few dominant voices, in addition to a broader spectrum of waves generated by turbulent fluid motion~\cite{Morozova_2018}.

The traditional {\tt BayesWave} algorithm models signals and noise transients as sums of Morlet-Gabor wavelets. There is currently no restriction on the overall amplitude and phase of the resulting waveform. In particular, neither are constrained to be smoothly varying. In contrast, {\tt BayesWaveVoices} ensures continuity of both the ampltide and phase by modeling the amplitude, $A(t)$, and instantaneous frequency, $f(t)$, of each ``voice'' as smooth functions such as splines or Legendre polynomials. The idea of modeling generic waveforms using splines to describe the amplitude and frequency evolution is not new~\cite{PhysRevD.96.102008,Mohanty_2023}, but our implementation of the model is very different as it uses trans-dimensional Bayesian sampling. Fixed dimension spline models have also been used to model possible departures from the predictions of general relativity~\cite{Edelman2020ConstrainingUP}.

We begin in Sec.~\ref{sec:improvements} by motivating the new approach by comparing the waveform reconstructions of the \BW and \BWV models using binary black hole mergers as an example. The \BWV model, basis functions, methodology and refinements to the model in the time domain, along with corresponding priors and proposals, are described in Sec.~\ref{sec:bwalgo}. Sec.~\ref{sec:C&D} discusses the results and conclusions drawn from the methodology presented in this paper and also concludes with a discussion of potential avenues for future development.

\section{Motivation}\label{sec:improvements}

The \BW algorithm models burst-type signals while accounting for any non-stationary and non-Gaussian detector noise features. The wavelet based \BW signal model works well for short duration signals and short duration noise transients~\cite{Cornish_2015, Littenberg_2015}. The new \BWV model is designed to cover longer duration signals.

Binary black hole mergers provide a good test bed for comparing the wavelet and voices models as the time that the signal spends in the sensitive band of the detectors can be adjusted by changing the total mass of the system.

To illustrate the performance of the wavelet and voices versions of {\tt BayesWave}, we consider simulated signals embedded in Gaussian noise with a spectrum that follows the advanced LIGO design sensitivity curve~\cite{aligo}. In contrast to analyses that use theoretical waveform templates, which allow for the model to be extrapolated beyond regions with significant signal power, generic models are only expected to recover the signal in the regions with significant signal power. 

\begin{figure}
\centering
\includegraphics[width=\columnwidth]{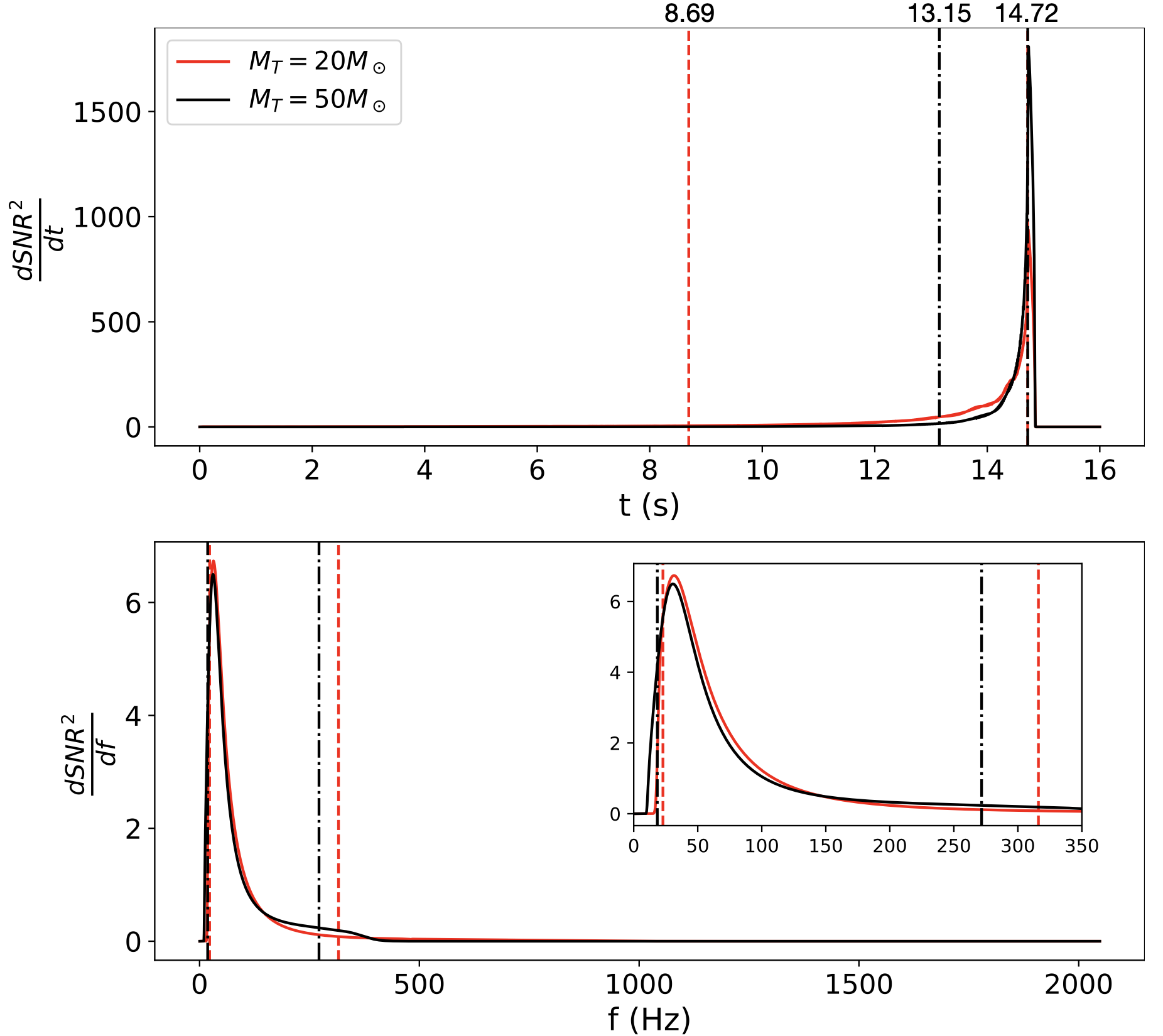}
\caption{\label{fig:dSNR2} Change in accumulated SNR$^2$ with time (upper) and frequency (lower) for a binary black hole system with total SNR = 20 and $M_T$ = 20$M_\odot$ (red line) and $M_T$ = 50$M_\odot$ (black line). The vertical lines bound the region that contains 90\% SNR$^2$.  We don't expect the signal reconstructions to have support outside of these regions.} 
\end{figure}

To illustrate the performance of the two models, we simulated 16 seconds of advanced LIGO data (for a single detector). In one example we added the signal from a binary black hole system with total mass $20 M_\odot$, and in the other we added a signal with total mass $50 M_\odot$. We used the IMRPhenomD waveform model, which describes the dominant harmonic for spin-aligned quasi-circular binaries~\cite{PhysRevD.93.044007}. The mass ratio and spins were $q = 1$, $\chi_1 = 0.2$, $\chi_2 = 0.1$, and the merger time was set at $t = 14.73 s$. Both signals were scaled to a total SNR of 20. Figure ~\ref{fig:dSNR2} illustrates the accumulation of SNR$^2$ over time, $d{\rm SNR^2}/dt$ and frequency $d{\rm SNR^2}/df$ for these examples. The vertical lines bound the region that contains 90\% SNR$^2$.
For a 20$M_\odot$ binary, these bounds, indicated by the red dashed line, cover a time span 6.02 seconds between 8.69s and 14.72s in time (top panel), and frequency bandwidth 293.3 Hertz between 22.8 Hz and 316.1 Hz in frequency (bottom panel). Similarly, for a 50$M_\odot$ binary the black dashed-dotted line represent the boundary of the sensitive region covers a time span of 1.57 seconds between 13.15s and 14.72s and a frequency bandwidth of 253.3 Hertz between 18.4Hz and 271.7Hz.  As expected, high mass systems spend less time in band than low mass systems.

\begin{figure*}
\centering
\includegraphics[width=2.0\columnwidth]{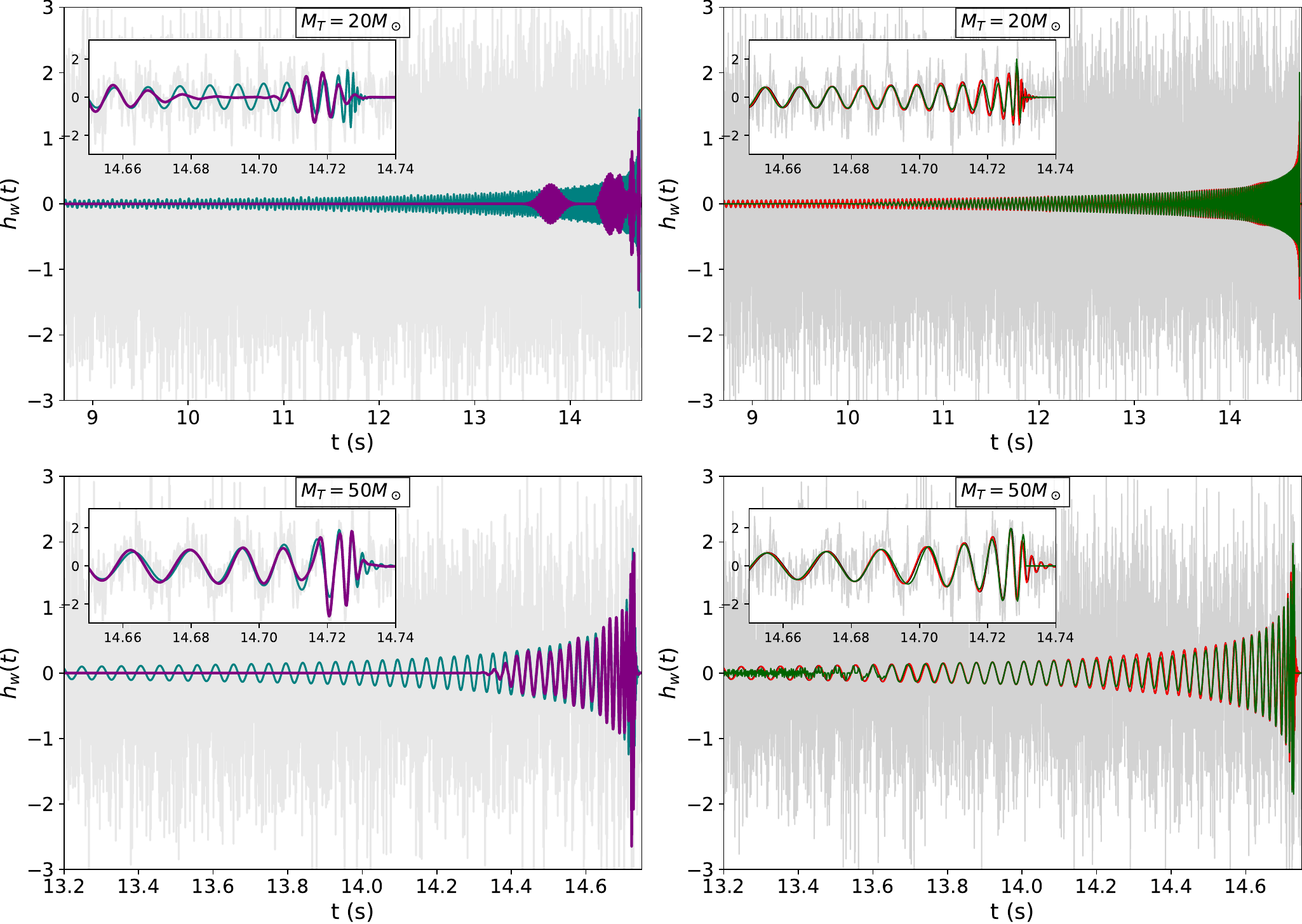}
\caption{ \BW~\cite{Cornish_2015, Cornish_2021} (left panel) and \BWV (right panel) reconstruction for an injected signal of SNR = 20. The injected signal is shown in teal in the left panels and in red in the right panels. The \BW reconstructions are shown in purple and the \BWV reconstructions are shown in green. The \BWV reconstructions use an Akima spline model that incorporates all refinements detailed in the subsequent sections of this paper. These plots are shown in the same 90\% SNR$^2$ region as indicated in Figure~\ref{fig:dSNR2}. The subplots show the zoomed in regions near merger. The top and bottom panels represent an equal mass binary of total mass 20$M\odot$ and 50$M_\odot$ respectively. Injected signal shown in teal and red a has spin of $\chi_1$ = 0.2 and $\chi_2$ = 0.1. The recovered median \BW reconstructions have a match of 0.567 and 0.823 for a 20$M\odot$ and 50$M_\odot$ respectively. Importantly, \BWV enhances the overall fit for the same binary system, achieving match values of 0.92 (for 20$M_\odot$) and 0.95 (for 50$M_\odot$).}
\label{fig:BW_BWV_compare}
\end{figure*}

Figure~\ref{fig:BW_BWV_compare} shows a comparison between the \BW (left panel) and \BWV (right panel) analyses for a simulated binary black hole signal of $M_T$ = 20$M\odot$ (top) and 50$M_\odot$ (bottom) with an injected SNR = 20. The simulated whitened data for a spin aligned black hole binary with spins $\chi_1$ = 0.2 and $\chi_2$ = 0.1 is shown in light grey. The time range shown in this Figure correspond to the 90\% SNR$^2$ regions shown in Figure~\ref{fig:dSNR2}. We observe that \BW encounters difficulties in capturing the signal for low-mass binaries ($M_T$ = 20$M\odot$). In contrast, \BWV effectively captures the high SNR signal region, achieving a match of 0.92, as opposed to \BW's match of 0.56. The improvement persists for the $M_T$ = 50$M\odot$ binary, with the \BWV reconstruction achieving a match of 0.95 compared to 0.82 for \BW, with the \BWV reconstruction extending to earlier times. The subplots in each panel highlight the high SNR region around merger. 

The match is computed using a point estimate for the waveform given by the median of the posterior samples of the waveform model. Here the match  between waveforms $h$ and $\bar{h}$ is defined as
\begin{equation}
    M = \frac{(h|\bar{h})}{(\bar{h}|\bar{h}) {(h|h)}} \, .
\end{equation}
where the inner product is defined such that $(a|b)=a^i C^{-1}_{ij} b^j$, where $C_{ij}$ is the noise covariance matrix.

\begin{figure}
\centering
\includegraphics[width=1.0\columnwidth]{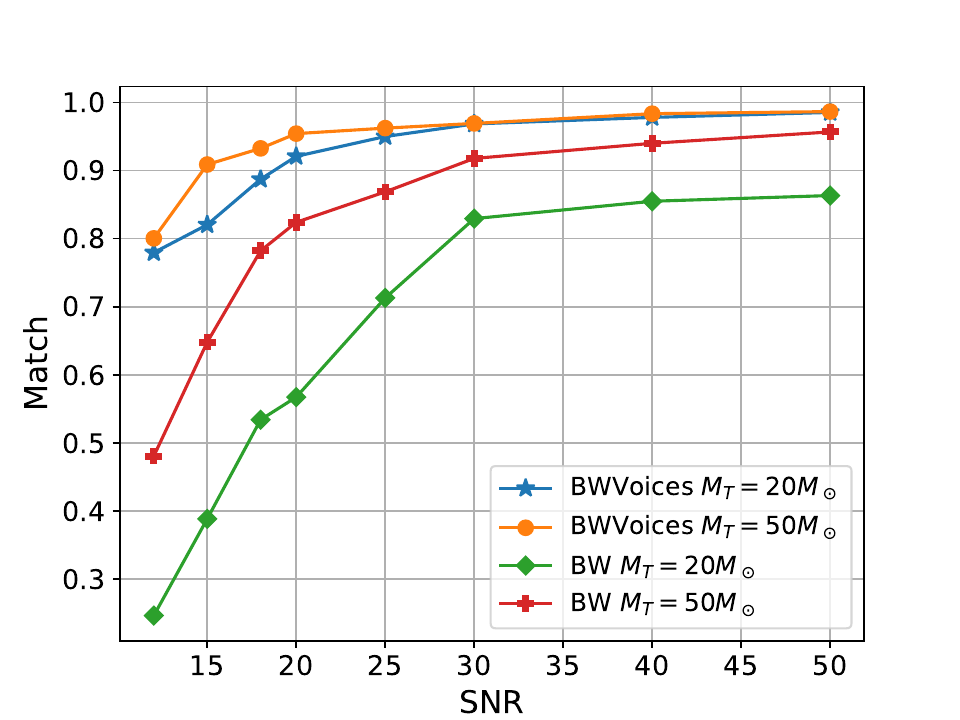}
\caption{ Matches between the simulated and recovered waveforms using \BW and \BWV as a function of signal-to-noise ratio for simulated black hole mergers. The green and red line shows \BW match for 20$M\odot$ and 50$M_\odot$ binary, while the blue and orange lines show the corresponding matches found using \BWV. The voices model significantly outperforms the wavelet model in all cases.}
\label{fig:snr_match_final_compare}
\end{figure}

The match as a function of SNR is shown in Figure~\ref{fig:snr_match_final_compare} for binaries with total mass 20$M\odot$ and 50$M\odot$ for both \BW and \BWV. The match increases with increasing SNR and $M_T$. The value approaches 1 for high mass and high SNR signals in either model. We notice that the \BWV model outperforms the \BW model for the entire SNR-mass parameter space. Note that the matches for \BW found here are somewhat lower than those shown in Ref.~\cite{Becsy:2016ofp}, the reason being that here we are using data at advanced LIGO design sensitivity, which results in longer time in band compared to the earlier work which used the lower sensitivities corresponding to the first observing runs.


\section{BayesWaveVoices Algorithm} \label{sec:bwalgo}

The \BWV algorithm models signals as a collection of ``voices'', which are each described by a smoothly changing amplitude and frequency. The current \BW algorithm computes the likelihood in the frequency domain, since the noise correlation matrix is diagonal in that representation, assuming that the detector noise is stationary over the relatively short time segments being analyzed. It might seem natural then to model the waveforms directly in the frequency domain such that for a single voice
\begin{equation}\label{hf_eqn}
    h(f) = A(f)e^{-i 2 \pi (\int^f {t(f') df'}) + \phi_0},
\end{equation}
One problem with this approach is that there is nothing stopping time from becoming multi-valued. This pathology is actually present in the frequency domain IMRPhenomD model~\cite{PhysRevD.93.044007}, which models $A(f)$ and the phase $\Phi(f)$. Extracting $t(f)$ via
\begin{equation}
    t(f) = \frac{1}{2\pi} \frac{d \Phi(f)}{df}
\end{equation}
reveals that time becomes multi-valued near merger for this model.

To avoid this pathology we instead model the signals in the time domain:
\begin{equation}\label{ht_eqn}
    h(t) = A(t)e^{-i 2 \pi (\int^t {f(t') dt')} + \phi_0}.
\end{equation}
There are several advantages to this approach. Firstly, time is single valued, and the model is able to accommodate signals that have instantaneous frequencies that can increase or decrease in time. Secondly, spectrograms of the data, such as the Q-scans used in \BW $tfQ$ proposal for placing wavelets, can be used to identify tracks in the time-frequency plane that serve as proposals for placing points along the frequency track $f(t)$. Thirdly, expressing the signal in terms of $A(t)$ and $f(t)$ allows for very efficient, automatically heterodyned likelihood calculations in the discrete wavelet domain~\cite{Cornish_2020}.

In the current implementation we chose to compute the likelihoods in the time domain to avoid having to Fourier transform the signals. The time domain analysis is performed using whitened data.
The data can be whiten either by transforming to the frequency domain, dividing the data by the amplitude spectral density $\sqrt{S(f)}$ then returning to the time domain, or by staying in the time domain and transforming the data using a Cholesky decomposition of the inverse of the noise correlation matrix. The signals can be whiten by the replacement $A(t) \rightarrow A(t)/\sqrt{S(f(t))}$. Alternatively, we can model the whitened amplitude directly.

\subsection{Basis Functions / Voices}\label{ssec:voices}

    Morlet-Gabor wavelets were originally chosen when constructing the \BW algorithm since they are maximally compact in time-frequency, and have simple analytic forms in both the time and frequency domains. But wavelets are not an essential part of the \BW algorithm. Far more important is the trans-dimensional modeling component, which can in principle build a signal out of any complete set of basis functions or over-complete frame functions. The performance of the algorithm can be improved if the chosen basis is able to capture signals using fewer parameters. For gravitational wave signals originating from coherent flows of matter and energy, we expect that the signal can be modelled by a few modulated voices, with distinct instantaneous frequencies. This is true for (low eccentricity) compact binary mergers~\cite{Moore_2018}, and appears to be a decent approximation for some core-collapse supernovae models~\cite{Morozova_2018} and for neutron star post-merger signals~\cite{Clark_2016}.
    
    These considerations motivate a new signal model based on the decomposition in time domain:
    \begin{equation}
        h(t) = \sum_{k=0}^{N_v} A_k(t) \cos{\left(\int^t 2 \pi f_k (t',\phi) dt' +\phi_k\right)},
    \end{equation}
    where $k$ labels each of the $N_v$ voices and $\phi_k$ is the overall phase for $k^{th}$ voice.
    
    There are many ways we can specify the functions describing the amplitude and frequency. For example, we could use a polynomial expansion (similar to the analytic post-Newtonian waveforms), or a sum of orthogonal basis functions (e.g. Chebyshev or Legendre polynomials), or a spline (as is used in the BW spectral model and for marginalizing over the LIGO calibration uncertainties), or some non-parametric model such as a Gaussian process. Regardless of the choice, the key idea is that the complexity of the model (e.g. the number of terms to keep in a polynomial expansion or the number of spline control points or knots to retain) should be determined from the data using trans-dimensional Bayesian inference. This allows the model to adapt and capture the inherent complexity of the gravitational wave signals present in the observed data. 

We chose to use Akima splines~\cite{Akima1970ANM} to model the smoothly evolving amplitude and frequency functions in time. Akima splines have compact support, and they are much less prone to develop unphysical oscillations than cubic splines. The location of the spline knots, their amplitude and also the number of spline knots are variable. These parameters are explored using a trans-dimensional Reversible Jump Markov Chain Monte Carlo (RJMCMC) algorithm. 

\subsection{Methodology}\label{ssec:method}

The \BWV algorithm returns samples from the posterior distribution of the waveform model. The waveform model is described by a collection of parameters $\th$ that specify the values and locations of the spline knots for the amplitude and instantaneous frequency of each voice. The dimension of $\th$ is variable. Bayes theorem
states that for a given set of data $\bold{d}$, the probability of measuring parameters $\th$ is expressed as 
\begin{equation}
    p(\th | \bold{d}) = \frac{p(\bold{d}|\th) p(\th)} {p(\bold{d})},
\end{equation}
where the posterior distribution function, $p(\th | \bold{d})$, quantifies the probability that the the model is described by parameters $\th$ given the observed data $\bold{d}$. The likelihood function, $p(\bold{d}|\th)$ or $\mathcal{L}$, evaluates the probability of observing data $\bold{d}$ given the parameter set $\th$. Assuming the noise is Gaussian distributed and the spectral model is fixed, the log likelihood can be written as
\begin{equation}
   \log  \mathcal{L} = -\frac{1}{2} \int_0^{T_{\rm obs}}(d_w(t) - h_w(t))^2 dt + {\rm const}\, .
\end{equation}
where the subscript $w$ indicates that the quantities are whitened. The prior probability distribution function, $p(\th)$, allows us modify the behavior of the spline model. The simplest choice, which we refer to as the reference model, assumes a uniform distribution for each parameter over some specified range.

The transdimensional MCMC method uses the Metropolis-Hastings algorithm~\cite{Metropolis1953EquationOS} to sample the parameter space. We start from an initial guess for $\th$ and then a new value is proposed based on the proposal distribution. The acceptance ratio, H, 
\begin{equation}
    H = \frac{ \mathcal{L}(\bold{d}|y) Q(x|y) P(x) } {\mathcal{L}(\bold{d}|x) Q(y|x) P(x)}
\end{equation}
is determined as the product of the likelihood ($\mathcal{L}$), the proposal distribution ($Q$), and the prior distribution ($P$). Here, $x$ represents to the current state and $y$ denotes the proposed state. The proposed jump is accepted if, $H>U(0,1)$, where $U(0,1)$ is a random number between 0 and 1, and rejected otherwise. This entire process is repeated until the model converges to a true posterior distribution.

In our analysis, we utilize Akima splines as the basis functions to model the time-varying amplitude $A_w(t)$ and frequency $f(t)$. We also explored different models, including splines~\cite{Ahlberg}, Steffen splines~\cite{Steffen1990ASM}, Legendre polynomials, and a smooth fitting function~\cite{article_smlinear}. Cubic splines, being global and $C^2$ differentiable, offer continuity and differentiability in the first and second derivatives. While this allows for flexibility, it can lead to undesired wiggles in the fitting functions. To address this, we introduce a separate prior on the derivatives of the cubic spline points to ensure their small values. However, this approach has the tendency to place control points too close to each other, resulting in oscillations in the fit over time. Steffen splines, on the other hand, exhibit high stiffness and demand monotonicity, presenting challenges in capturing complex features. Legendre polynomials, being a global basis functions defined between the interval 1 and -1, require additional scaling and expansion coefficients, introducing extra parameters.

A common characteristic of the aforementioned basis functions is their global nature, meaning that altering one point affects the entire curve. In contrast, Akima splines are more local in nature. They are evaluated over a five-point stencil and are $C^1$ differentiable, making them well-suited for our analysis. Due to their local nature and smoothness, Akima splines are more efficient in capturing localized features and are therefore chosen as the basis functions for modeling the waveform parameters throughout the rest of the paper.

In this model, the location of spline knots, their amplitudes, the coalescing phase $\phi_c$ and even the number of spline knots are variable. The key idea of using Akima splines voices with variable number of spline knots within the framework of trans dimensional MCMC is to avoid oscillations and facilitate a balance between model intricacy and the fidelity of fit. This approach embraces a natural parsimony, allowing the data to determine the delicate balance between model complexity and goodness of fit.

We inject simulated binary black hole signal generated from the IMRPhenomD phenomenological model~\cite{PhysRevD.93.044007}. This model accurately describes the $l = |m| = 2$ harmonic mode for a spin-aligned, non-precessing quasi-circular binary, incorporating information from numerical relativity. This waveform model provides amplitude and phase information in the frequency domain. The time domain amplitude $A(t)$ and phase $\phi(t)$ are extracted using the method described in section V.D of Ref.~\cite{Cornish_2020}. The instantaneous frequency is computed as $f(t) = \dot{\phi}/2 \pi$ and the whitened amplitude as $A_w(t) = A(t)/\sqrt{S(f(t)}$. 

 The trans dimensional Markov Chain Monte Carlo (RJMCMC) algorithm~\cite{10.1093/biomet/57.1.97, 23b6b6ec-42a0-3e0a-8bf2-dc861b85327c} is used to vary the number of spline knots in each model and to vary the values and locations of the knots.

In reality, we do not know in advance where to place the boundaries of the model in time, so we have to make the start and end times of the spline model variable (the alternative is to fix the first and last knots at the the end points of the time interval being analyzed, but this forces the model to cover regions where there is no information about the signal.)

In the current implementation we restrict the model to use a single voice ($N_v=1$). Going forward we plan to allow $N_v$ to vary as dictated by the data.

\subsubsection{Priors and Proposals}\label{sec:priors}

Priors on the number and placement of spline knots in the amplitude and frequency models can be used to enhance the performance of the algorithm. In what we refer to as the base model, the priors on all these quantities are taken to be uniform in some range. The prior on knot point locations follows a uniform distribution $U(0, T_{obs})$, where $T_{obs}$ represents the duration of the analysis segment, and we define $t=0$ to be the start time of the analysis segment. The magnitude of amplitude and frequency priors are $U(0, A_{\rm max})$ and $U(f_{\rm min}, f_{\rm max})$ respectively. We set $A_{\rm max}$ at some multiple of the noise variance $\sigma^2=1$, with a typical choice being $A_{\rm max}= 6 \sigma^2$. In the current study with simulated data we set $A_{\rm max}$ to twice the maximum amplitude of the injected signal. We set $f_{\rm max}$ equal to the Nyquist frequency corresponding to the given sample rate and $f_{\rm min}$ is set equal to 10Hz, reflecting the steep loss in sensitivity below this frequency. The prior distribution for the overall phase parameter is defined as a uniform distribution in the range $U(0,2 \pi)$ associated with amplitude and frequency is specified as  $U(0,N_{\rm max})$, where a typical choice of $N_{\rm max} = 150$ is an effective choice.


A variety of proposals are used to update the amplitude and frequency models. These include prior draws on the location and values of the spline knots. In the within-dimensional update, the amplitude is either proposed from a uniform distribution or a Gaussian jump centered around the current state. Lateral moves facilitate the movement of spline points in time, with the randomly selected location proposed as a Gaussian jump about its current position.

The location of the end knots is allowed to vary, so that the model can expand or shrink in time to cover the region where there is significant signal power. In the default model,
the splines for the amplitude and frequency share the same end knots. This adds some complications to the sampling. It is possible to allow the models to have different end knot locations, but then we either have to have some prescription for how to extend the models to cover the same total time interval, or we have to restrict the time interval to the region where the two models both have support.

Transdimensional moves involve adding a point anywhere between the current end-points. The knot locations are drawn uniformly between the boundaries, and amplitudes are drawn from the prior distribution. The end points are treated differently since they are shared between the two models. Jumps in either endpoint location are proposed from a Gaussian distribution centered on the current location. Jumps are rejected if the proposed location is inside any of the other knot locations.
Updates of the overall phase parameter $\phi_0$ are drawn from a Gaussian distribution centered on the current value.


Special jumps are permitted for amplitude and frequency models. Proposals involve drawing from a density centered around the fit, with reverse moves drawing a point near the current Akima fit and using the fit at {\it x} (the previous state) as a reference. Alongside these special proposals, we include symmetric proposals, such as a uniform draw from the prior range or a Gaussian jump from the current location with a selection of different variances. We ensure detailed balance and verify prior recovery for model reversibility. This comprehensive set of proposals enhances the exploration of the parameter space throughout the Bayesian inference process.

In principle, the model should be able to lock onto a signal starting from a random draw from the prior. However, Bayesian models, especially complex trans-dimensional models can take a long time to converge from a random starting point. A better approach is to start the model off with a good guess. For example, a scalorgram of the whitened data you be used to identify the region in time and frequency where a signal is present, and could also be used to extract a guess for the the amplitude and frequency tracks. In the current analysis we have focused on the sampling after a signal has been identified, and we have yet to develop a general purpose initialization algorithm. In this study we are focusing on simulated binary black hole mergers with known waveforms, so we are able to initialize the model using our knowledge of the injected waveforms.

We initialize our model by judiciously placing spline knots positioned in amplitude and frequency using the linear least squares fitting methods. The procedure involves  calculating the chi-squared of the linear fit until it surpasses a predefined tolerance, at which point a knot is placed. This process is repeated, proceeding from the established knot point to the subsequent one, until the chi-squared value once again exceeds the specified tolerance. This iterative placement of knots persists until all points are positioned up to $~$0.01 seconds after the merger event. Subsequently, the model places spline points at intervals of 0.5 seconds until reaching the end of the injected signal. The selection of the tolerance value is aimed to strike a balance. It is not set too high to prevent having too few points initially, nor is it set too low to avoid an excess of control points that may not be useful.


Consider two reference system with fixed masses $m_1$ = $m_2$ = 10$M_{\odot}$ and $m_1$ = $m_2$ = 25$M_{\odot}$ with spins $\chi_1$ = 0.2 and $\chi_2$ = 0.1. We set the initial model with NA and NF number of spline points, defining the time-varying $A_w(t)$ and $f(t)$ functions using the least square fitting method discussed above. Our reference model uses uniform priors and proposals for location, amplitude and number of spline knots for both amplitude and frequency model. The boundaries are variable. 

\subsection{Testing the model}\label{ssec:C&M}


\begin{figure}[h]
\centering
\includegraphics[width=\columnwidth]{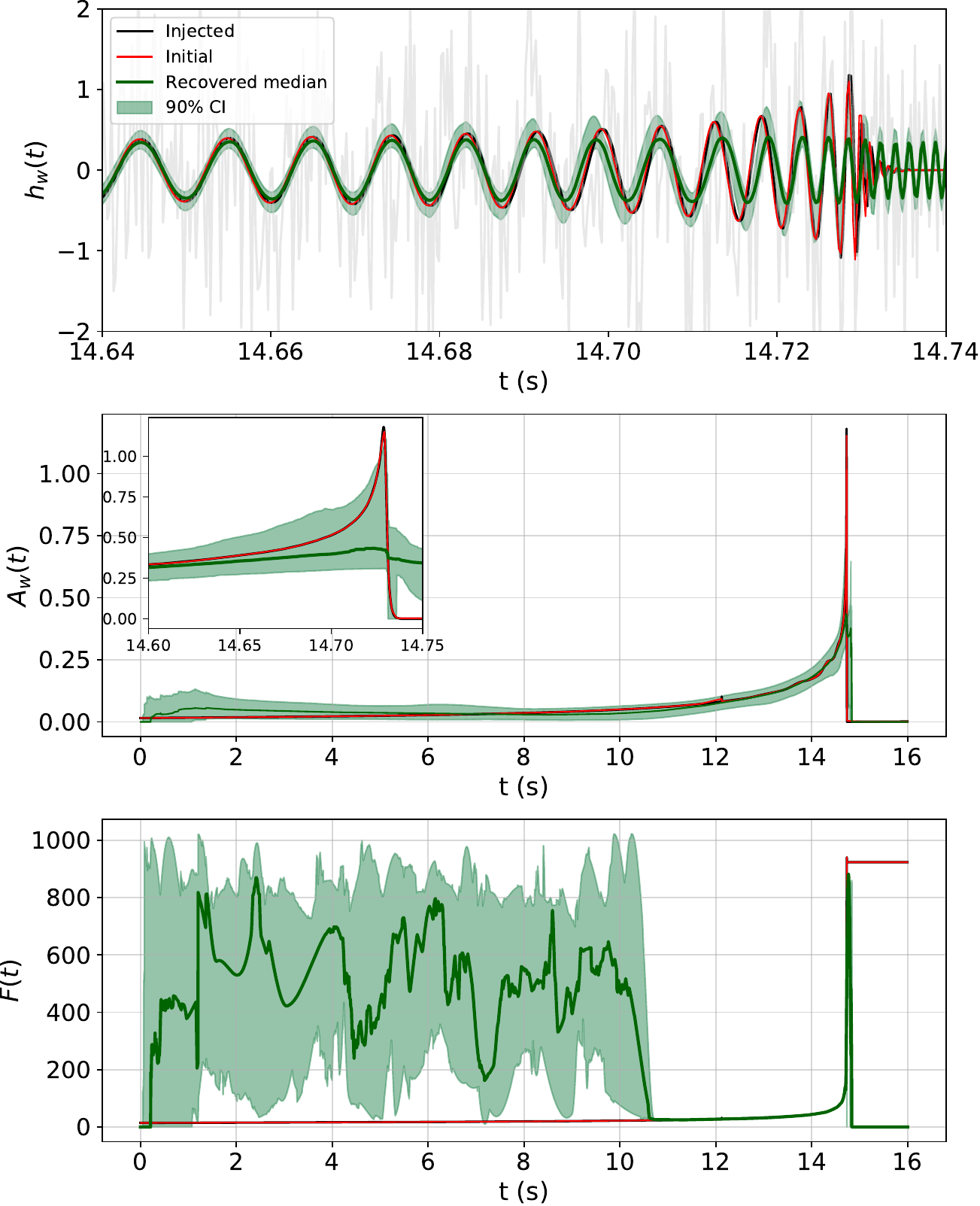}
\caption{The top panel shows the median waveform reconstruction in green with the 90\% credible interval using an Akima spline model with variable boundaries starting with the full time range of 16s. The injected (black line) signal is same as the initial start signal(red line) over the full length scale. The middle panel shows the injected amplitude in black which again is same as the initial start amplitude. A noteworthy observation is the abrupt decay in amplitude just after the merger, leading to suboptimal performance of the model. The lowest panel shows the reconstruction for frequency as a function of time.  The model encounters challenges in accurately fitting the initial times due to a low signal-to-Noise ratio during these early stages. This resulted in a match = 0.85 using median waveforms. }
\label{fig:ref}
\end{figure}


To test the \BWV algorithm, we simulated Gaussian noise with a spectrum that follows the advanced LIGO design curve, and injected signal of $M_T$ = 20$M_\odot$ and SNR = 15. For our initial test we start with our model defined in the entire time range as our boundaries, the boundaries are allowed to move based on the likelihood estimates as defined in the reference model. Figure~\ref{fig:ref} shows the reconstructed waveform (top panel) and its 90\% confidence interval (CI), while the injected and initial signals span the entire time range. These results are from our reference model defined in Sec.~\ref{sec:priors}.

Inspecting the amplitude extracted from the simulated signal, depicted in red in middle panel of Figure~\ref{fig:ref}, reveals a sharp feature at the merger. One of the key feature of splines is to enforce smoothness, and the Akima splines used in the analysis struggle to model sharp features in the amplitude or frequency evolution.

The frequency model, shown in the lower panel of Figure~\ref{fig:ref}, recovers the frequency track fairly well for roughly two seconds before merger, which is in keeping with our expectations that the model should recover the signal in regions with significant contribution to the SNR$^2$. However, contrary to expectations, the boundaries for the amplitude and frequency models did not shrink down to focus on the high SNR$^2$. The locations for the end knots did move around as the RJMCMC sampler explored the model space, but the end knots typically covered a much wider region than where the signal was concentrated. At early times, where there is little signal power, the frequency model covers most of its prior range, indicating that it was not recovering the signal in this region. Ideally the boundaries of the model would have shrunk to focus on the region where the signal power is concentrated.

The inability of the amplitude model to capture sharp features, and the fact that the boundaries did not shrink to focus on the region of high SNR$^2$ suggests that the base model can be improved. 

\begin{figure}[h]
\centering
\includegraphics[width=1.0\columnwidth]{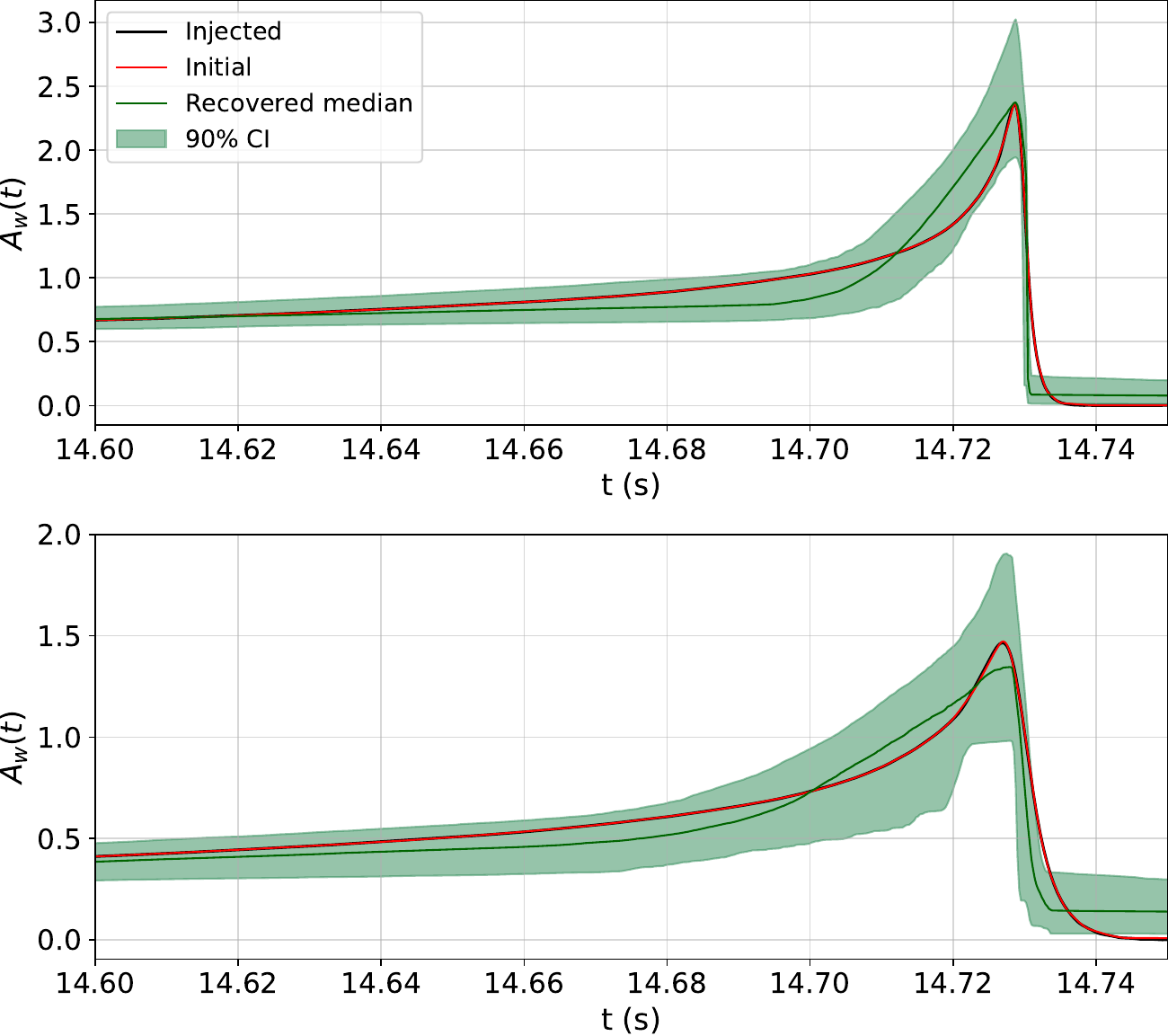}
\caption{Amplitude reconstruction depicted in green using the basic Akima spline model, with the initial signal shown in red covering the entire range, alongside the injected signal in black. The top panel displays the amplitude reconstruction for a system with SNR = 30 and $M_T = 20 M_{\odot}$, while the bottom panel showcases SNR = 15 and $M_T = 50 M_{\odot}$. Notably, the amplitude reconstruction effectively captures sharp features in signals with high mass and high SNR.}
\label{fig:amp}
\end{figure}

While the spline model may not readily capture sharp features, given sufficient SNR, it demonstrates an ability to do so. Additionally, for lower masses, the SNR is distributed, whereas for higher masses, a more concentrated SNR is observed. Figure~\ref{fig:amp} illustrates the behavior of Akima splines for a system with $M_T = 20 M_{\odot}$ and SNR = 30 (top panel) and another system with $M_T = 50 M_{\odot}$ and SNR = 15 (bottom panel). It is evident that increasing either the SNR or the mass compels Akima splines to adeptly handle sharp discontinuities. Therefore, it is a combined effect of total mass and SNR; with sufficient power, the model exhibits proficiency in fitting sharp features, extending its capability even to SNR = 20.

\begin{figure}
\centering
\includegraphics[width=1.0\columnwidth]{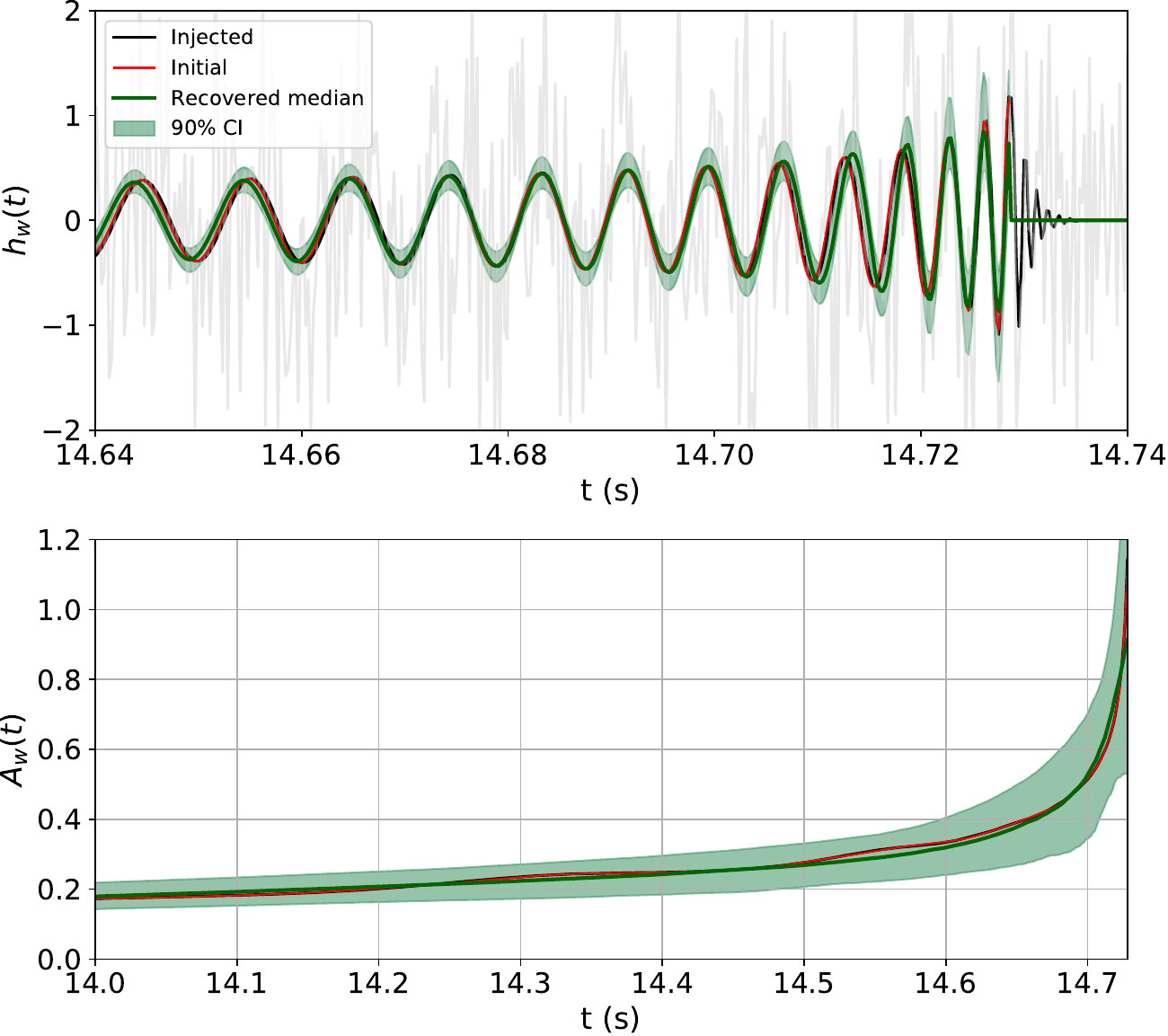}
\caption{ Median reconstruction is shown in green for a system with SNR = 15 and $M_T = 20M\odot$. A basic Akima spline model with fixed boundaries and an injected signal in black while an initial start signal covering 95\% SNR$^2$ region shown in red, which doesn't include the sharp drop in amplitude near merger as can be seen in the bottom panel. The model captures the peak around merger and recovers the signal with match = 0.88. The shaded region is the 90\% credible interval for the reconstruction.}
\label{fig:hbasicsnr95}
\end{figure}

To understand these issues better, we looked at what happens when the boundaries of the spline models are fixed to cover the time interval during which 95\% of the SNR$^2$ is deposited. Crucially, in this example, the 95\% region does not include the ringdown portion of the waveform, which we saw poses challenges for the reference \BWV model.
Figure ~\ref{fig:hbasicsnr95} shows the injected signal in black and the portion of the signal that accounts for 95\% of the SNR$^2$ in red. The median of the signal model posterior is shown in dark green, along with the 90\% CI in lighter green.  The match for the recovered median waveform over the fixed region is 0.88, which improves on the fit we found when the boundaries were free to move. In this fixed boundary example where the signal is cut off at merger, the amplitude smoothly increases with no sharp drop, and we see in the bottom panel that the spline model is able to accurately recover the peak amplitude of the signal, in contrast to the case where the amplitude drops sharply through ringdown.

\begin{figure}[h]
\centering
\includegraphics[width=1.0\columnwidth]{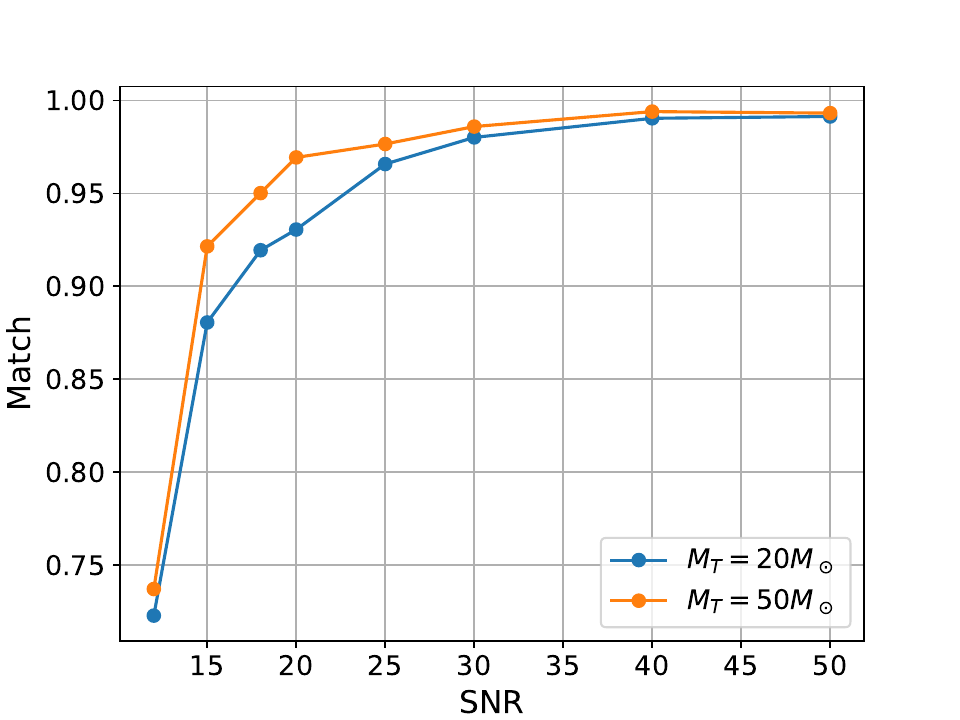}
\caption{Match as a function of signal-to-noise ratio using basic Akima spline model and fixed boundaries covering 95\% SNR$^2$ region for binary systems of 20$M\odot$ and 50$M_\odot$, excluding the sharp drop around the merger. The match increases with increase in SNR and mass and slowly asymptotes to unity.}
\label{fig:snrmatch_basic}
\end{figure}

The match versus SNR curve for this case is depicted in Figure~\ref{fig:snrmatch_basic} for a binary black hole signal with a total mass of 20$M_\odot$ and 50$M_\odot$. The graph suggests that as the SNR increases, the match proportionally rises, eventually stabilizing at a near-constant value close to unity. Hence, the model performs effectively for signals initiated with fixed boundaries and no abrupt discontinuities.

However, it is essential to enable variable boundaries in reality, as we lack information on where to place the boundaries in time. Allowing the locations of the end knots in the amplitude and frequency models to vary adds some complications to the sampling, especially if both models are required to share the same end points. The advantage is that the \BWV model inherits some of the advantages of the wavelet version since the model is then compact in time and frequency.

\begin{figure}[h]
\centering
\includegraphics[width=0.9\columnwidth]{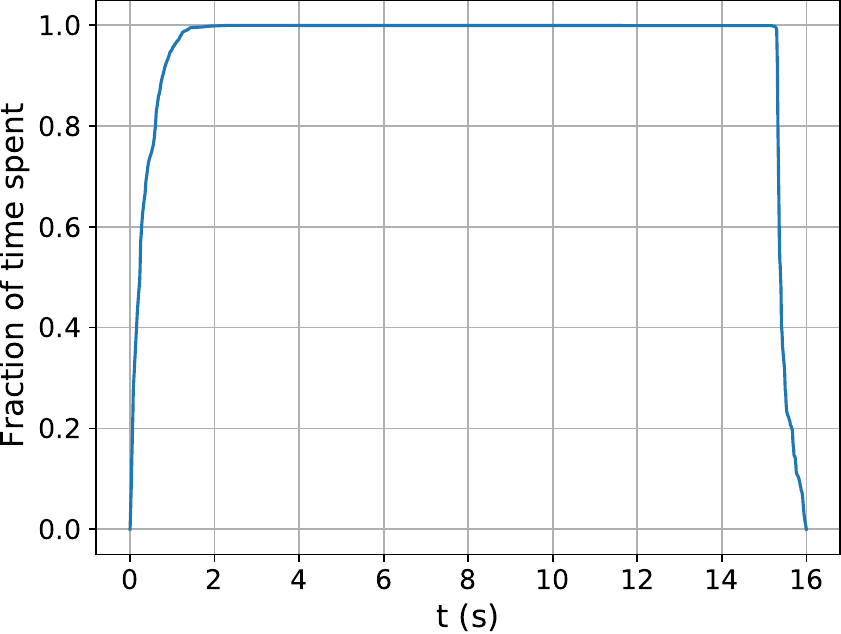}
\caption{Fraction of active time for a specific temporal region in the basic model using Akima splines with variable boundaries, as generated through transdimensional MCMC iterations.}
\label{fig:fraction}
\end{figure}

To enhance the understanding of the variable boundary, Figure~\ref{fig:fraction} illustrates the fraction of time a specific region remains active. As the boundaries shift, the spline is modeled within that particular temporal segment. It is observed that the model exhibits reduced preference for about initial 2 s region but becomes active thereafter, extending up to the merger. Essentially, the end boundary retracts to exclude the post-merger region where SNR is very low.

\begin{figure}
\centering
\includegraphics[width=0.9\columnwidth]{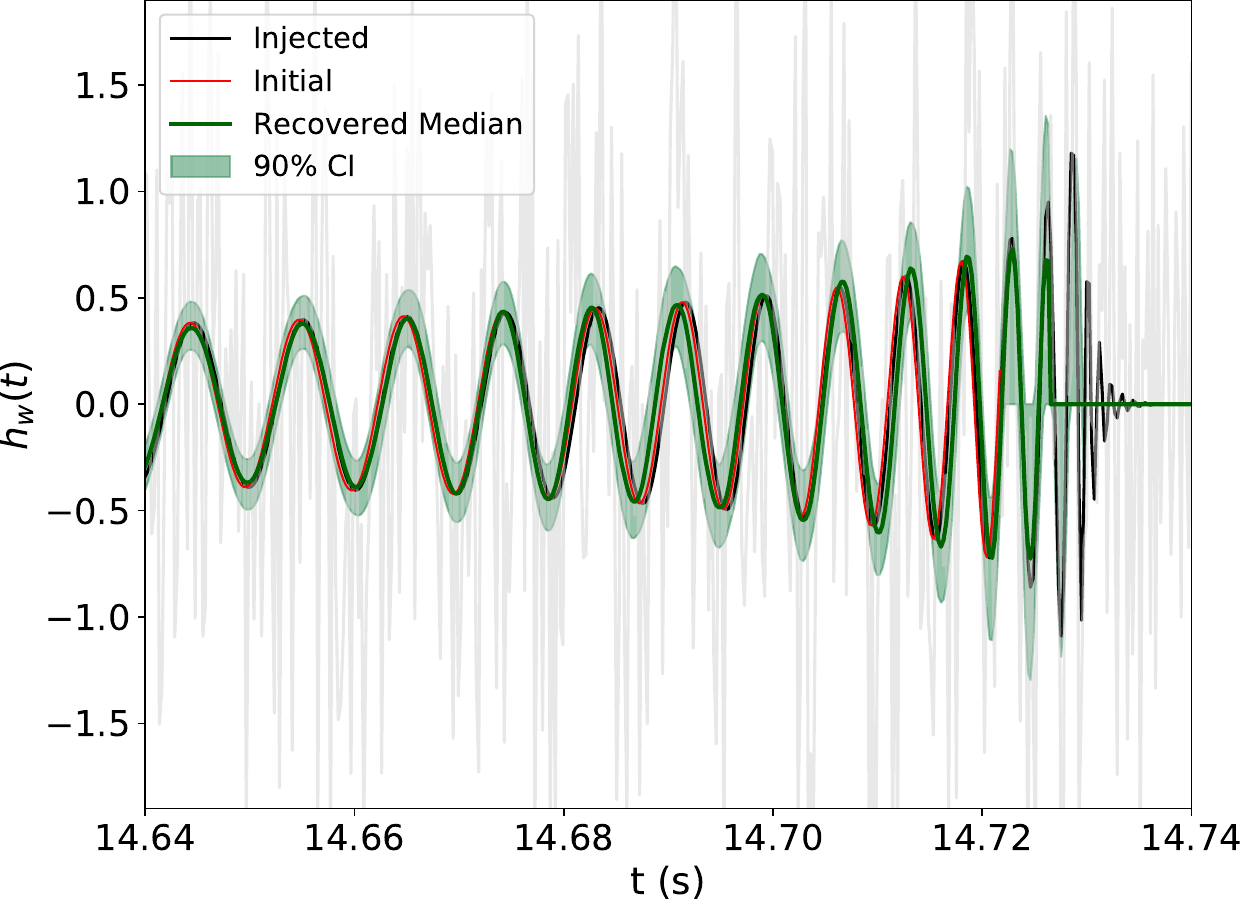}
\caption{The injected signal is shown in black while the signal is initiated with the end boundary at 14.724s (red line), about 6ms before merger. The boundaries are allowed to move and model extends its end boundary location to 14.728s extrapolating the information to about 2ms before merger. This reconstruction recovers signal of SNR = 15 and $M_T = 20M_\odot$ with a match of 0.88. }
\label{fig:hx_vb}
\end{figure}

The inclusion of additional information and complexities necessitates an enhanced model focused on capturing the most interesting and informative regions. If we consider continuous $A_w(t)$ and $f(t)$ without any discontinuity and allow the boundaries to move, the defined model should extend to effectively capture the merger region. Figure~\ref{fig:hx_vb} illustrates how the recovered signal not only captures the initial signal but also extends beyond to capture the injected features for the same reference case discussed above. This replicates the effects that an original full-length signal would exhibit. The end boundary, initially at t = 14.724s, moved to 14.728s while capturing the high SNR region and extending to encompass the merger region with an overall match of 0.88. This improvement in the model allows for a more accurate representation of the gravitational wave signal, showcasing its flexibility in adapting to diverse signal characteristics and optimizing information retrieval.

The presence of sharp features in initial model significantly impacts the overal fit. In scenarios where amplitude is the sole parameter, decoupled from frequency, Akima splines excel in capturing these sharp discontinuities. Extensive testing has confirmed the effectiveness of this approach. However, the challenge arises when frequency and amplitude are coupled in likelihood estimations, introducing added complexity to the modeling process. The interdependence of these parameters necessitates a more sophisticated approach, considering their joint influence on the likelihood calculations.

\subsection{Further refinements}\label{refine}

\begin{figure}
\centering
\includegraphics[width=1.0\columnwidth]{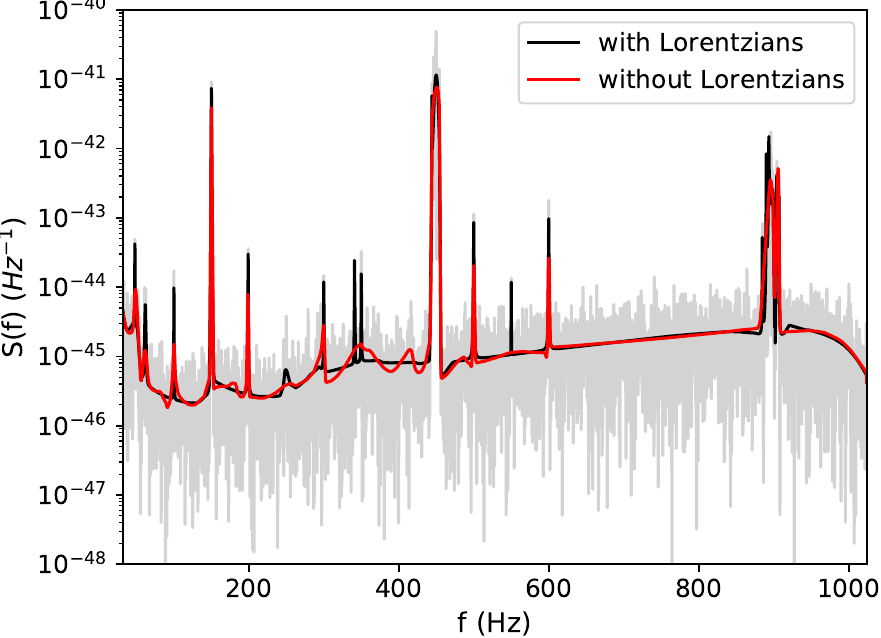}
\caption{PSD estimation for a 4s Virgo data with a 2048Hz sample rate is illustrated. In light gray, a periodogram of the data is presented, while the spectral estimates using a \BW model are shown in red (utilizing only the smooth component of the PSD model) and in black (including the Lorentzian line feature modeling). The model with the Lorentzian feature more accurately captures the sharp features compared to the ones without.}
\label{fig:spec}
\end{figure}

By their very nature, spline models have a tendency to smooth over sharp features. To overcome this limitation, the model can be extended to include sharp features in addition to the smooth spline model. This is what is done in the {\tt BayesLine} Bayesian spectral estimation algorithm~\cite{Littenberg_2015,Gupta:2023jrn}, where a combination of splines and sharp line features, modeled as Lorentzians, are used to model the LIGO/Virgo noise power spectra. An example of the {\tt BayesLine} spectral modeling is shown in Figure~\ref{fig:spec}, one with just the spline model and the other with splines and lines. The model without Lorentzian lines misses many of the sharp, low-amplitude (low SNR) line features. The spline model is able to pick up the high amplitude (high SNR) lines, showing that with enough SNR the splines can fit any shape, despite their preference to produce smooth curves. The model that includes lines does a much better job of fitting the power spectrum (as measured by the Anderson-Darling statistic computed for the whitened data~\cite{Gupta:2023jrn}.

 The effectiveness of the multi-component {\tt BayesLine} model motivated us to introduce a similar strategy by incorporating a variety of additional functions that are better suited to modeling sharp features. We included five distinct types of growing and decaying functions, all appropriately scaled to yield similar curves for a comparable rate parameter, to help ensure that proposals that seek to to swap one function for another will get accepted. These tapering functions can either be 
 rising or falling, and are scaled to go from 0 to 1. The complete model for $A_w(t)$ is given by the product of the spline model and the tapers.
 
\begin{figure}
\centering
\includegraphics[width=1.0\columnwidth]{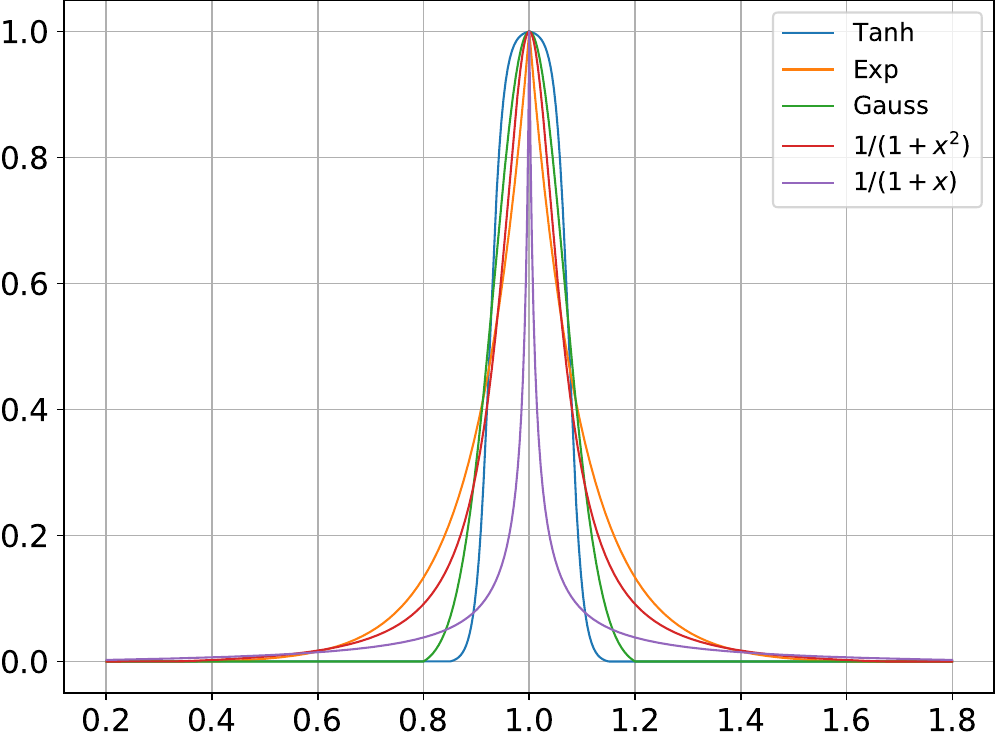}
\caption{ Five distinct spike functions, all appropriately scaled to yield similar curves are shown. The purple, green and blue line represents a tan hyperbolic, exponential and Gaussian function respectively. The orange and yellow curve shows inverse power law function of $1+x^2$ and $1+x$ respectively. Smooth Akima splines can be multiplied by (1 + spike) to potentially incorporate the sharp changes in amplitude for a given signal.}
\label{fig:spike}
\end{figure}

We extend this approach by incorporating a more refined strategy, employing a combination of both growing and decaying functions to construct a spike feature. Figure \ref{fig:spike} illustrates five distinct functions utilized to create this spike. This method goes beyond a simplistic swap between functions, allowing for different components on the left and right sides of the peak.
For instance, the spike could comprise a Gaussian fit on the left side and a power law on the right side of the peak. This flexibility enables the model to adapt to various signal features, capturing intricate details that a singular function might overlook. Specifically, for a binary black hole merger, a spike that rises gradually slowly and then decays rapidly (exponentially) should fit the final part of the amplitude fairly well.
To enhance the fitting process, we started by dividing the injected amplitude by this spike function. Subsequently, by fitting the initial spline model to this modified amplitude, the product of the spline and the spike is expected to yield a robust initial fit with faster convergence. This approach not only allows for the seamless transition between different functions but also enables tailored adjustments on either side of the peak, enhancing the model's adaptability and performance. Although this additional feature moderately improves the overall fit, further refinements were required to enhance signal reconstruction.

To ensure that splines exhibit a smooth fall-off and avoid unexpected oscillations, we introduce a mechanism involving predefined knots in a specific time region. Let $\Delta T_s$ and $\Delta T_e$ represent the time region where this predefined function $\mathcal{S}(t)$ is evaluated. The locations of the fixed knots are positioned at some fraction of $\Delta T_s$ and $\Delta T_e$, with their amplitudes gradually approaching zero on either ends.
The spline function takes into account these fixed knot data points during evaluation, ensuring a smooth fall-off on both ends. Although the spline is evaluated within the specified boundary (as discussed in Sec.~\ref{sec:C&D}), the inclusion of these additional knots is incorporate to help mitigate oscillations, especially in regions of low SNR, as observed in the initial times in the frequency fit (see bottom panel Figure \ref{fig:ref}).

It is worth noting that splines are not inherently local, meaning the positions of control points influence a broader region. By guiding the amplitudes to gracefully approach zero, we enhance the spline model's behavior, as abrupt endings are generally unfavorable for splines. This approach prioritizes a smooth descent to zero, contributing to the model's stability and reliability in capturing complex signal characteristics.
Mathematically, the predefined function for the two ends can be expressed as follows:
\begin{align}
    \mathcal{S}_{start}(t) = A_w(t) \frac{e^{-(t + \Delta T_s - T_s)} -1 }{e^1 - 1}, \\ 
    \mathcal{S}_{end}(t) = A_w(t) \frac{e^{-(t + \Delta T_e - T_e)} -1 }{e^{-1} - 1},
\end{align}
where $\mathcal{S}_{start}(t)$ and $\mathcal{S}_{end}(t)$ indicated the two predefined functions before the start and after the end boundary knots.

To prevent frequency from exhibiting unwanted oscillations during the MCMC sampling process, a potential solution is to initiate the frequency integration from the peak of the maximum amplitude. Cumulative integration is specifically centered around the frequency point at which the maximum amplitude occurs. This approach aims to avoid the accumulation of unwanted area at low SNR, particularly given the limited SNR in the initial times.

As we allow the boundaries to move for capturing the signal around the maximum power, we have the option to either tie both the frequency and amplitude boundaries to move together or allow them to move independently. Untying these boundaries provides the model with flexibility in defining frequency and amplitude where needed, potentially reducing the total number of parameters to work with. Although untied boundaries increase the complexity of the model, but they offer greater flexibility. We chose to have the model set the amplitude to zero outside of the amplitude model end knots and to set the frequency to a constant outside of the frequency model end knots.

It's important to note that the first and last points with tied boundaries remain fixed, meaning there are no birth or death moves; they only move when there is a boundary update. Consequently, the endpoints do not have a uniform prior. To address this, we initially run the model with a constant likelihood, where the prior for the first and last points is exponentially growing and decaying, respectively. The prior is then fitted to a function, and its inverse is used as a counter-prior. This counter-prior, obtained from the inverse of the prior in the constant likelihood run on the first and last points with proper normalization, mitigates the effect of the non-uniform prior on the location of knot points. 

In an ideal scenario, the amplitude would serve as a guide for the model to allocate more points in regions where the signal undergoes rapid evolution. If the amplitude is properly normalized to the noise level, the probability of placing points is influenced by the amplitude plus a random variable drawn from a uniform distribution U(0,1). Consequently, in regions with higher amplitude, the prior probability of placing points is elevated.

Building on this concept, we introduced a prior based on the second derivative of the amplitude given by
\begin{equation}
    \left(\frac{dA_w}{dt}\right)^2 + \left|A_w\frac{d^2A_w}{dt^2}\right|.
\end{equation}
This addition is motivated by the fact that this prior scales with SNR$^2$, representing the curvature of the SNR$^2$. While it might seem redundant given the inherent influence of likelihood on knot placement, this prior becomes particularly valuable for low SNR signals. Physically, gravitational wave signals are not random segments of power in time and frequency but rather exhibit clustering. Hence, we incorporate a model for the cluster prior, which is particularly informative for low SNR signals.

Although the likelihood should guide the model to place knots where sharp features are present, the physical model benefits from explicit guidance regarding regions where the signal undergoes rapid variation. To achieve this, we normalize $d^{2}|A_w^2|$ to unity and create a probability function. This function is then utilized as a prior draw, with a 50\% weightage to strike a balance and avoid overly stringent priors that would only place knots where derivatives are high. This nuanced approach aims to enhance the adaptability of the model while respecting physical constraints and data characteristics.

After exploring the introduced modifications, the signal reconstruction closely resembles the top right panel of Figure~\ref{fig:BW_BWV_compare}. For a binary black hole signal with $M_T = 20M_\odot$ and SNR = 20, the  match is 0.92. Furthermore, for an even lower SNR = 15 injected signal with the same remaining parameters, the match is 0.87.

\begin{figure}
\centering
\includegraphics[width=1.0\columnwidth]{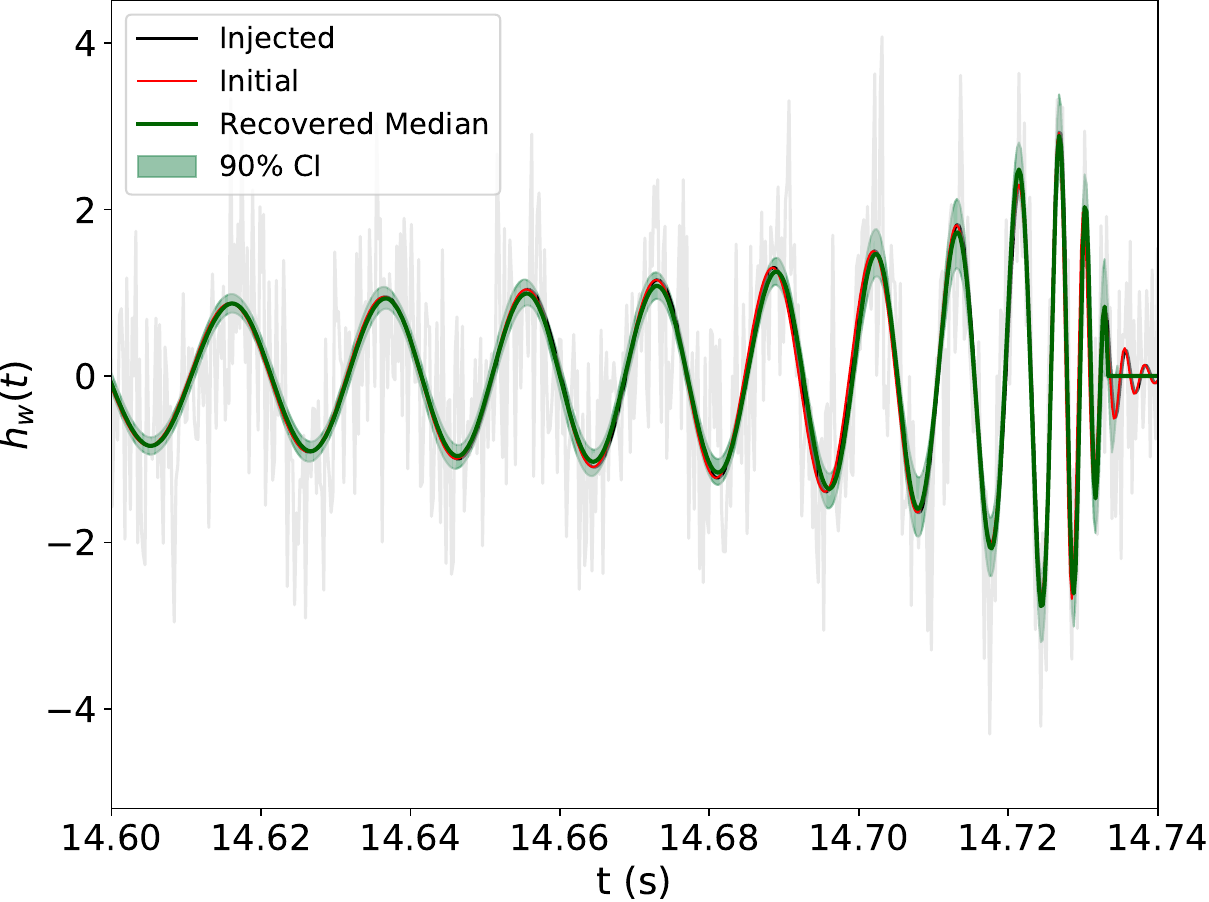}
\caption{For a high mass binary of $M_T = 50M_\odot$ and SNR = 30, the recovered reconstruction (green) of the injected signal (black) with sharp features yield a match of 0.97 using the current \BWV model. This model does not use the derivative prior on amplitude showing a more realistic case.}
\label{fig:hfsnr30_m50}
\end{figure}

Considering a slightly more massive and higher SNR signal, the \BWV reconstruction is presented in Figure~\ref{fig:hfsnr30_m50}, achieving a match of 0.97. Additionally, Figure~\ref{fig:snr_match_final_compare} illustrates a general trend where massive systems tend to exhibit higher matches than lower mass systems. This phenomenon arises from the relatively higher signal amplitude compared to noise in the case of a massive binary. It can be asserted that \BWV consistently outperforms \BW for signals of varying masses, demonstrating effectiveness for both weak and strong signals. Nonetheless, there remains untapped potential for further enhancement. This potential could be realized through the implementation of more focused proposals or the incorporation of a well-defined spike model. Such refinements hold promise for enabling the model to effectively capture sharp features, especially in scenarios characterized by low SNR.

\section{Conclusions and Future Direction}\label{sec:C&D}

We conclude that a localized spline model like Akima splines are suited to avoid oscillatory behaviors of cubic and monotonicity of Steffan splines. Although with the imposed priors and proposals, and added refinements to the model, the current \BWV model as it stands now is not just an Akima fit but with multiple layers of added constraints. 

Our analysis sets the stage for further exploration by incorporating multiple detectors and including information about both cross and plus polarizations. Extending the \BWV model could provide invaluable insights into the most intriguing potential signals from core collapse supernovae and conducting neutron star post-merger studies. The associated gravitational radiation is poised to provide abundant information about the underlying dynamics and processes driving supernovae. Representing any signal as a product of smoothly evolving amplitude and phase over time, \BWV offers a promising avenue for studying these signals all the way up to the merger, particularly in neutron star post-merger scenarios.

Furthermore, the algorithm we have created opens new avenues for testing the general theory of relativity. By introducing a correction $\Delta h(t)$ to the waveform model and constraining the parameters of these perturbations using gravitational wave and binary pulsar observations, we can explore and validate the predictions of general relativity in the context of gravitational wave phenomena.

\section{Acknowledgment}

This material is based upon work supported by NSF's LIGO Laboratory which is a major facility fully funded by the National Science Foundation. The authors are grateful for the support provided by NSF award PHY 2207970.

\newpage
\section*{References}

\bibliography{bibliography}

\end{document}